\documentclass[aps,prl,twocolumn,superscriptaddress,showpacs]{revtex4-1}

\usepackage{graphicx}%
\usepackage{color}
\usepackage{epstopdf}
\usepackage{amssymb}
\usepackage{amsmath}
\usepackage{amsfonts}

\begin{document}
\title{Magneto-optical probe of the fully gapped Dirac band in ZrSiS}

\author{E. Uykur}
\email{ece.uykur@pi1.physik.uni-stuttgart.de}
\affiliation{1.~Physikalisches Institut, Universit{\"a}t Stuttgart,
70569 Stuttgart, Germany}

\author{L. Z. Maulana}
\affiliation{1.~Physikalisches Institut, Universit{\"a}t Stuttgart,
70569 Stuttgart, Germany}

\author{L. M. Schoop}
\affiliation{Department of Chemistry, Princeton University, Princeton, NJ, 08544, USA}

\author{B. V. Lotsch}
\affiliation{Max Planck Institute for Solid State Research, 70569 Stuttgart, Germany}

\affiliation{Chemistry Department, University of Munich (LMU), 81377 M{\"u}nchen, Germany}

\author{M. Dressel}
\affiliation{1.~Physikalisches Institut, Universit{\"a}t Stuttgart,
70569 Stuttgart, Germany}

\author{A. V. Pronin}
\affiliation{1.~Physikalisches Institut, Universit{\"a}t Stuttgart,
70569 Stuttgart, Germany}

\date{October 17, 2019}

\begin{abstract}

We present a far-infrared magneto-optical study of the gapped
nodal-line semimetal ZrSiS in magnetic fields $B$ up to 7~T. The
observed field-dependent features, which represent intra- (cyclotron
resonance) and interband transitions, develop as $\sqrt{B}$ in
increasing field and can be consistently explained within a simple
2D Dirac band model with a gap of 26~meV and an averaged Fermi
velocity of $3\times10^{5}$~m/s. This indicates a rather narrow
distribution of these parameters along the in-plane portions of the
nodal line in the Brillouin zone. A field-induced feature with an
energy position that does not depend on $B$ is also detected in the
spectra. Possible origins of this feature are discussed.
\end{abstract}

\maketitle

Three-dimensional nodal-line semimetals (NLSMs) \cite{Burkov2011}
are currently a subject of intensive experimental
investigations~\cite{Fang2014, Bian2016, Yu2015, Chen2015, Xie2015}.
In these systems, 2D electronic bands possessing linear dispersion
cross each other along continuous lines in reciprocal space, hence
enabling 2D Dirac electrons to exist in the 3D bulk. Among these
materials, ZrSiS is probably the best studied
system~\cite{Schoop2016, Chen2017, Neupane2016, Ali2016,
Pezzini2017, Matusiak2017, Hu2017, Schilling2017}. This material
possesses two types of nodal lines: one of them is situated far away
($> 0.5$ eV) from the Fermi level E$_{F}$, while the second one,
most interesting, is much closer to E$_{F}$ and is fully gapped due
to spin-orbit coupling. This gap is calculated to be very small, of
the order of 10 meV \cite{Schoop2016}. Angle-resolved photoemission
spectroscopy (ARPES) provides values between 15 and 50
meV~\cite{Chen2017}, while our earlier optical studies estimate the
maximum of the gap to be around 30 meV~\cite{Schilling2017}.
Although the low-energy nodal line in ZrSiS is slightly gapped, the
linearity of the electronic bands forming this line extends up to
$\sim$0.5~eV, and other non-linear bands do not cross E$_{F}$. This
makes ZrSiS one of the best systems to study the properties of Dirac
electrons in NLSMs.

Magneto-optical spectroscopy is a powerful tool to investigate
electronic properties of nodal semimetals and narrow-band-gap
materials~\cite{Orlita2011, Orlita2012, Chen2017a, Chen2014a,
Shao2019}. For example, this method enables experimental
verification of the electronic band structure by tracing the optical
transitions between the magnetic-field-induced Landau levels (LLs).
This approach is particularly relevant at low energies, where other
experimental techniques often lack accuracy and resolution. Here, we
report on far-infrared magneto-optical investigations of ZrSiS. We
found that Dirac quasiparticles fully dominate the ac
(magneto)transport in ZrSiS: both inter- and intra-band optical
transitions demonstrate a square-root dependence on magnetic field,
typical for such bands. All measured magneto-optical spectra can be
well described by a simple model of gapped Dirac bands with a single
(i.e. $\mathbf{k}$-independent) gap of 25 meV.

The sample studied was grown by the method described elsewhere
\cite{Schoop2016}, and was the same single crystal previously used
for the zero-field optical measurements of
Ref.~\onlinecite{Schilling2017}. The optical reflectivity spectra
were collected from (001) surfaces (the in-plane response) utilizing
a home-build magneto-optical setup connected to a commercial
Fourier-transform infrared spectrometer. The spectra were recorded
for photon energies between $\sim 5$ and $75$ meV in magnetic fields
up to 7~T at 10~K. The Voigt geometry was chosen as the measurement
configuration to allow the Kramers-Kronig analysis for linearly
polarized light~\cite{Palik1970} (in the commonly utilized Faraday
geometry, such analysis would only be meaningful for circular
polarizations, which cannot be used for broadband optical
measurements). The spectra have been obtained with two linear
polarizations, $\mathbf{\tilde{E}||B}$ and $\mathbf{\tilde{E}\perp
B}$ (here, $\mathbf{\tilde{E}}$ is the electric component of the
probing radiation and $\mathbf{B}$ is the external magnetic field).
Hereafter we describe our results obtained for
$\mathbf{\tilde{E}\perp B}$, while in the $\mathbf{\tilde{E}||B}$
polarization no field-induced changes were detected. A simultaneous
change of the $\mathbf{B}$-field direction and the light
polarization (keeping the angle between them fixed) in respect to
the crystallographic direction within the (001) plane did not affect
the spectra due to the tetragonal crystal symmetry within this
plane. In order to obtain the optical conductivity, the measured
reflectivity spectra were merged with the zero-field reflectivity at
higher energies (up to $\sim 5.6$ eV, \cite{Schilling2017}) and the
Kramers-Kronig analysis was performed using x-ray atomic scattering
functions~\cite{Tanner2015} and Drude-Lorentz fits as extrapolations
at high and low frequencies, respectively.

%%%%%%%%%%%%%%%%%%%%%%%%%%%%%%%%%%%%%%%%%%%%%%%%%%%%%%%%%%%%%%%%%%%%%%%%%%%%%%
\begin{figure}[ht]
\centering
\includegraphics[width=\linewidth]{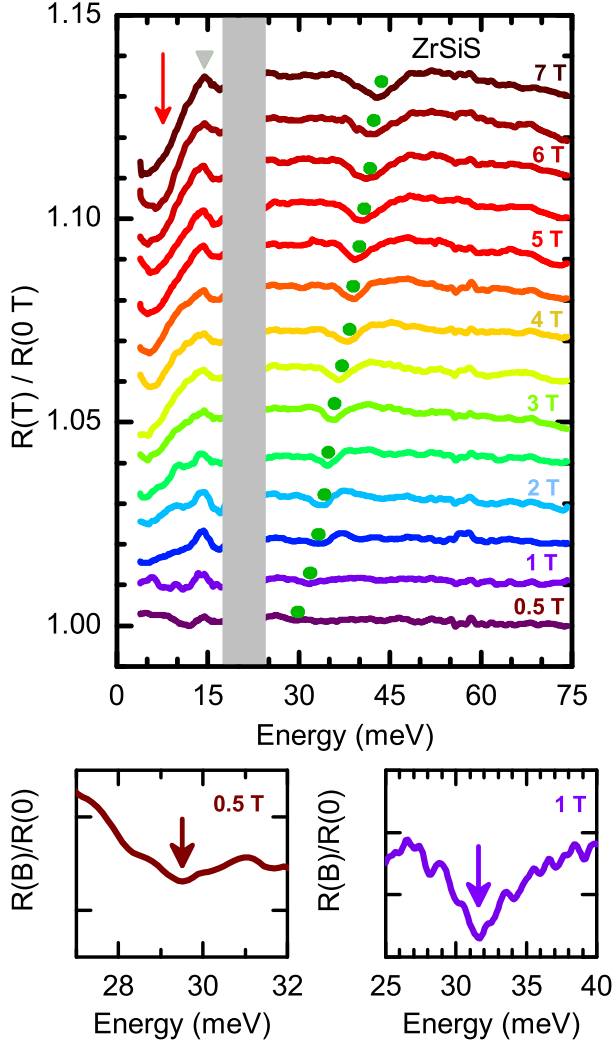}%
\caption{Magneto-reflectance spectra of ZrSiS normalized to
zero-field reflectivity. The spectra at various fields are shifted
by 0.01 with respect to the previous one for clarity. The shaded
area is the range not accessible for our measurement setup. Enlarged
spectra near one of the absorption modes for 0.5~T and 1~T are given
in the bottom panels.}
\label{Ref}%
\end{figure}
%%%%%%%%%%%%%%%%%%%%%%%%%%%%%%%%%%%%%%%%%%%%%%%%%%%%%%%%%%%%%%%%%%%%%%%%%%%%%%

\begin{figure*}[ht]
\centering
\includegraphics[width=1\linewidth]{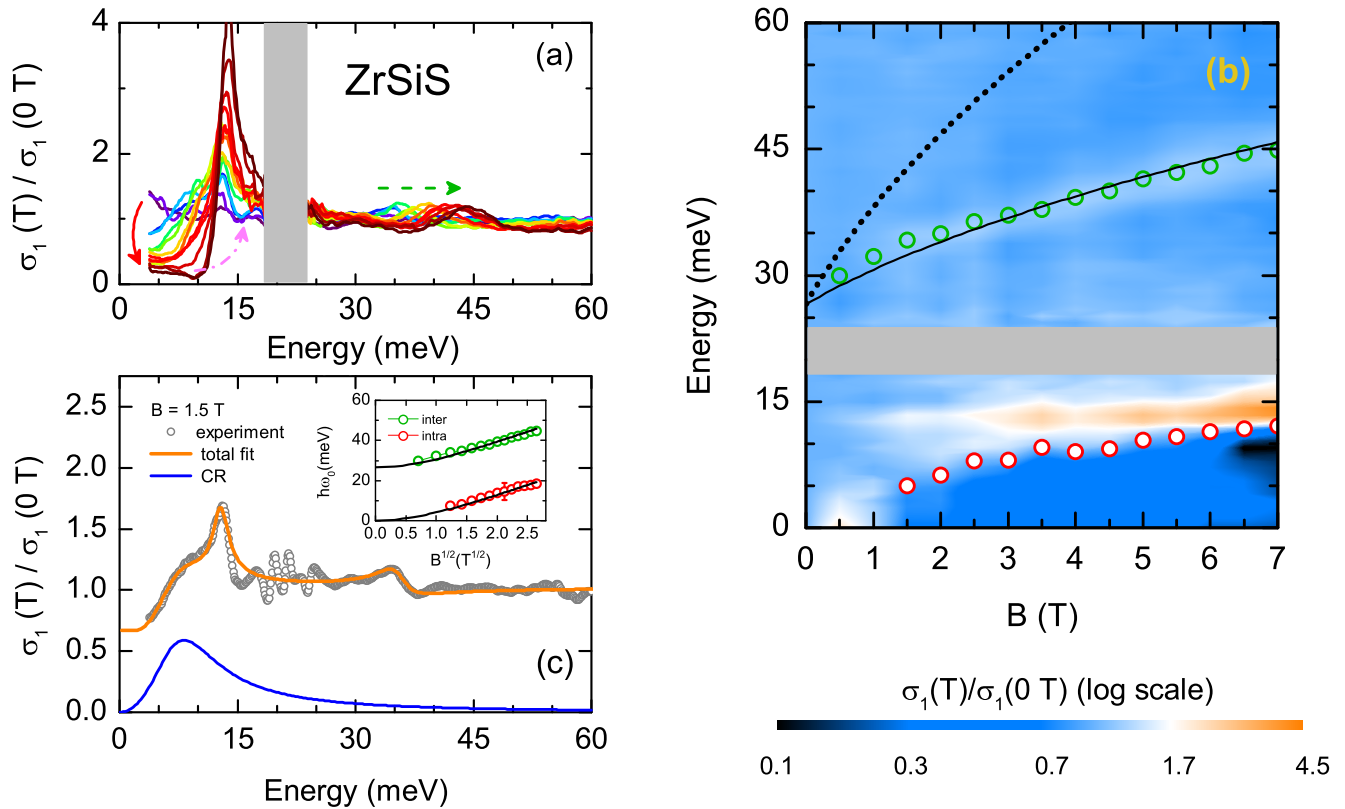}%
\caption{(a) Relative optical conductivity of ZrSiS \textit{vs.}
frequency for all studied fields. (b) Same data as a false-color
plot. Lines and dots correspond to different inter-LL transitions as
discussed in the text. (c) An example of the Lorentz-Drude fit to
the optical conductivity data at 1.5 T. The structures between
approximately 17 and 24 meV are noise (corresponding to the shaded
areas in the other panels). The inset shows the energy positions
($\hbar\omega_{0}$) of the inter- and intra-band transitions
determined from the conductivity data and fitted with
Eq.~(\ref{LLtransitions}). Note the $\sqrt{B}$ horizontal scale.}
\label{contour}%
\end{figure*}

The measured reflectivity is given in Fig.~\ref{Ref}. To demonstrate
the $B$-field-induced changes more clearly, we plot the relative
reflectivity, $R(B)/R(B = 0)$. The corresponding spectra of the real
part of optical conductivity, normalized by its zero-field value,
$\sigma(B)/\sigma(B = 0)$, are given in Fig.~\ref{contour}(a). The
spectra between $\sim 17$ and $24$~meV are largely affected by noise
in our setup and therefore eliminated from the discussion. Above
$50$ meV no field-induced features appear; we chose the energy scale
on the plots accordingly.

%%%%%%%%%%%%%%%%%%%%%%%%%%%%%%%%%%%%%%%%%%%%%%%%%%%%%%%%%%%%%%%%%%%%%%%%%%%%%%
\begin{figure}[t]
\centering
\includegraphics[width=7 cm]{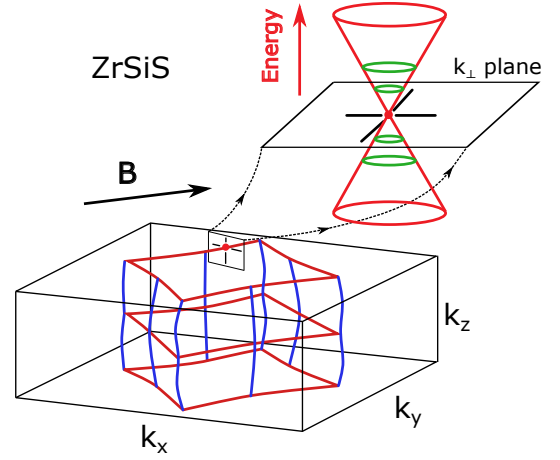}%
\caption{Schematics of the nodal line in ZrSiS. The 2D Dirac band
with its Landau levels formed for the in-plane orientation of
\textbf{B} is sketched in the upper right corner ($k_{\bot}$ stands
for the $k$-directions perpendicular to the nodal line). The band
gap is not shown for simplicity.}
\label{nodal_line}%
\end{figure}
%%%%%%%%%%%%%%%%%%%%%%%%%%%%%%%%%%%%%%%%%%%%%%%%%%%%%%%%%%%%%%%%%%%%%%%%%%%%%%

The optical spectra demonstrate three field-induced features: (i) a
suppression of reflectivity and optical conductivity at the lowest
frequencies (marked with the red arrows in both figures); (ii) a
feature that grows in intensity with increasing $B$, but remains its
frequency position constant at around 15~meV [the gray triangle in
Fig.~\ref{Ref} and the strongest peak in Fig.~\ref{contour}(a)]; and
finally, (iii) a feature around 30-45 meV seen at all the fields and
marked with the green circles in Fig.~\ref{Ref} and the green arrow
in Fig~\ref{contour}(a) demonstrating the frequency shift of the
feature as $B$ increases. In Fig.~\ref{contour}(b), a false-color
plot of the relative optical conductivity as a function of energy
and magnetic field is shown. The three field-induced features are
seen in this plot as changes of color. The features (i) and (iii)
are further emphasized by the open symbols obtained as discussed
below.

%%%%%%%%%%%%%%%%%%%%%%%%%%%%%%%%%%%%%%%%%%%%%%%%%%%%%%%%%%%%%%%%%%%%%%%%%%%%%%
%%%%%%%%%%%%%%%%%%%%%%%%%%%%%%%%%%%%%%%%%%%%%%%%%%%%%%%%%%%%%%%%%%%%%%%%%%%%%%

\begin{figure}[t]
\centering
\includegraphics[width=1\linewidth]{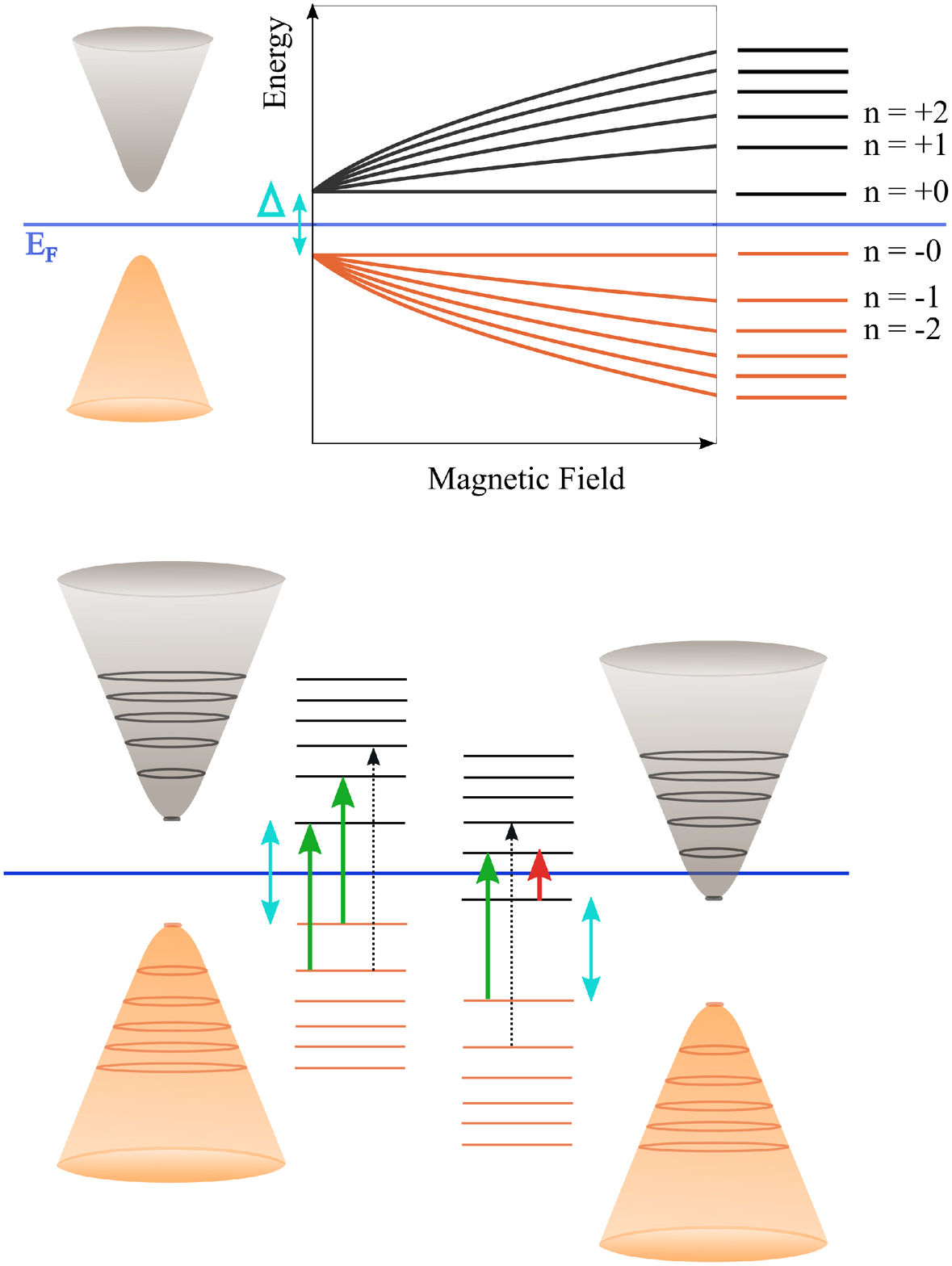}%
\vspace{-0.5 cm}%
\caption{Upper panel: Schematic representation of
Landau levels appearing in a gapped Dirac band. Bottom panels sketch
the 2D Dirac band in ZrSiS. Two cuts perpendicular to the nodal line
are shown: along some portions of the gapped nodal line, the Fermi
level is in the gap (left picture), while in the other portions
(right picture) it is in the conduction band. The cyan solid arrow
represents the determined single band gap. Other solid arrows depict
the observed transitions: the intra-band (CR) transition $+0
\rightarrow +1$ is shown as a red arrow, while the interband
transitions ($-0 \rightarrow +1$ and $-1 \rightarrow +0$) are given
as green arrows. The dotted arrows show other possible interband
transitions, which are not observed in this work. The color code for
the arrows is consistent with the one used to mark the transition
modes in Figs.~\ref{Ref} and \ref{contour}.}
\label{sketch}%
\end{figure}

Before going to the discussion, let us recall that for gapped Dirac
bands the LL spectrum reads as~\cite{Chen2014a, Chen2014b}:
\begin{equation}
E_{\pm n} = \pm \sqrt{2e\hbar\mid n \mid B v_F^2+\left(\frac{\Delta}{2}\right)^2},
\label{LL}
\end{equation}
where $n$ is the LL index, $v_F$ is the (average) Fermi velocity,
and $\Delta$ is the gap between conduction and valence bands. The
LLs in the conduction and valence bands correspond to the $``+"$ and
$``-"$ signs in Eq.~(\ref{LL}), respectively. The spectrum of
Eq.~(\ref{LL}) is shown in the upper panel of Fig.~\ref{sketch}. For
the allowed inter-LL transitions, $\mid n \mid$ can be changed by
$\pm 1$. Hence, the transition energy can be written, e.g., as:
\begin{eqnarray}
E_t = \sqrt{2e\hbar(\mid n \mid + 1) B
v_F^2+\left(\frac{\Delta}{2}\right)^2} \nonumber \\ \pm
\sqrt{2e\hbar\mid n \mid B v_F^2+\left(\frac{\Delta}{2}\right)^2}
\label{LLtransitions}
\end{eqnarray}
with the plus sign corresponding to interband and the minus sign to
intraband transitions. These transitions are schematically depicted
in the bottom part of Fig.~\ref{sketch}. Two situations are
considered: with E$_F$ in the gap (left panel) and in the conduction
band (right panel). Both situations are relevant for ZrSiS: it is
well known that the Fermi level crosses electronic bands along some
portions of the nodal line, while it is within the gap
elsewhere~\cite{Schoop2016}. In the following, we argue that this
simple model can account for all \textit{field-dependent} features
observed by us in the magneto-optical spectra of ZrSiS.

It is worth noting here that optical-conductivity measurements are
not $k$-sensitive and probe the entire Brillouin zone. In the
present study, we mostly probe the in-plane parts of the nodal line,
because the external magnetic field is applied parallel to the (001)
plane and hence the closed cyclotron orbits are formed for the
carriers from the bands forming the in-plane portions of the nodal
line, see Fig.~\ref{nodal_line}. (Small contributions from the
out-of-plane parts can also be present due to their corrugations.)

We start from the lowest energy feature (i). To make it more
visible, we added the red open circles to Fig.~\ref{contour}(b).
Their positions are determined as the unity-crossing points of
$\sigma(B)/\sigma(B = 0)$, cf. Fig.~\ref{contour}(a). The feature
(i) starts to appear at 1.5~T and seems to extrapolate to zero as
$B\rightarrow 0$. By looking on the field evolution of this feature
in Fig.~\ref{contour}(a), it can clearly be related to depletion of
the Drude response: the loss of Drude spectral weight at low
energies is expected to be transferred to a cyclotron-resonance (CR)
mode~\cite{Orlita2012}, as indicated by the dash-dotted pink arrow
in Fig.~\ref{contour}(a). In our measurements the cyclotron
resonance is superimposed on the field-independent feature (ii). In
order to extract the CR parameters, we performed a Drude-Lorentz
fit~\cite{Dressel2002} that takes into account the CR itself, the
leftover Drude component, the field-independent resonance, and the
high-energy mode corresponding to feature (iii), the latter is taken
into account only for completeness. In Fig.~\ref{contour}(c), we
present an example of such a fit for 1.5~T. This way we can extract
the field dependence of the cyclotron resonance. In the inset of the
Fig.~\ref{contour}(c) we plot the obtained energies of the cyclotron
resonance as a function of $B$ field. The linear behavior on the
$\sqrt{B}$-scale proves the relativistic nature (linear bands) of
the carriers responsible for this resonance and can be nicely fitted
by Eq.~(\ref{LLtransitions}) using $n=0$ and the minus sign
corresponding to the intraband transition (the red arrow in
Fig.~\ref{sketch})

Now, we turn to the feature (iii). At low $B$, it extrapolates to
the gap detected in our zero-field measurements~\cite{Schilling2017}
and it can be assigned to the transitions between LLs in the valence
and conduction bands. In Fig.~\ref{contour}(b), we plot the peak
positions of $\sigma(B)/\sigma(B = 0)$ as open green symbols.
Remarkably, we can fit the symbols with Eq.~(\ref{LLtransitions})
using the same $\Delta$ and $v_{F}$ as we used to fit the CR mode.
Only the sign between the addends should be changed (minus to plus).
Thus, the observed mode (iii) corresponds to the interband
transitions involving the $n=0$ LLs, i.e., the transitions from
$n=-1$ to $n=+0$ or from $n=-0$ to $n=+1$ (green arrows in
Fig.~\ref{sketch}).

The fits to the B-dependent modes (i) and (iii) shown in
Fig.~\ref{contour}(b,c) as black solid lines are all performed using
the single values of $\Delta = 26$ meV and $v_{F} = 3\times10^{5}$
m/s. This strongly indicates that the distribution of these
parameters along (the in-plane portions of) the nodal line in the
Brillouin zone is rather narrow, as optical probes may only provide
momentum-averaged quantities. Had $\Delta$ and $v_{F}$ possessed
strong $\mathbf{k}$-dispersions, the field-induced magneto-optical
features would be extremely broad, if detectable. As already
noticed, the gap identified is well within the margins provided by
calculations~\cite{Schoop2016}, ARPES~\cite{Chen2017}, and
optics~\cite{Schilling2017}. The same is true for the obtained Fermi
velocity, cf. Refs.~\cite{Ali2016, Lodge2017}.

In Fig.~\ref{contour}(b), we also show the position expected for the
inter-LL transitions from $n=-2$ to $n=+1$ and from $n=-1$ to $n=+2$
(dotted line, see also dotted arrows in Fig.~\ref{sketch}). These,
as well as other transitions between the LLs with larger $n$, are
not detectable in our spectra. This situation is quite common: the
line broadening increases with energy~\cite{Orlita2011,
Nedoliuk2019} making the line detection difficult.

%%%%%%%%%%%%%%%%%%%%%%%%%%%%%%%%%%%%%%%%%%%%%%%%%%%%%%%%%%%%%%%%%%%%%%%%%%%%%%
%%%%%%%%%%%%%%%%%%%%%%%%%%%%%%%%%%%%%%%%%%%%%%%%%%%%%%%%%%%%%%%%%%%%%%%%%%%%%%

Finally let us turn to the field-independent feature (ii), which is
seen around 15~meV. The position of this mode does not show any
appreciable frequency shift, but the mode strength increases with
magnetic field. A simple explanation of this mode would be an in-gap
impurity resonance. However, the studied ZrSiS sample is very clean:
the mean free path extracted from our optical measurements on this
very sample is of the order of 1~$\mu$m at low
temperatures~\cite{Schilling2017}. Thus, this explanation seems not
to be very likely. Alternatively, surface states~\cite{Topp2017}
might be relevant for the formation of this mode. A more appealing
option is of bulk and intrinsic origin. We note that a low-energy
mode with a very similar behavior in magnetic field (constant energy
position and strength increasing with $B$) was observed, but
remained unexplained, in another gapped NLSM,
NbAs$_{2}$~\cite{Shao2019}. It would be tempting to assign both
modes to the B-independent resonance, recently theoretically
proposed to be a hallmark of the NLSM state~\cite{Duan2019}.
However, the model used in this reference (a single nodal loop
formed by crossing cones) is far too simple and not directly
applicable to ZrSiS or NbAs$_{2}$: both compounds possess complex
and extended nodal lines. Thus, we call for more theoretical studies
in this direction.

Summarizing, our in-plane far-infrared magneto-optical study of
ZrSiS reveals three features developing in applied magnetic field.
The energy position of two of them scales as a square root of $B$. A
simple model of a single gapped Dirac band provides adequate
explanation for the field evolution of these features, the
low-energy feature being associated with intra-band (cyclotron
resonance) absorption, while the high-energy one appearing due to
interband LL transitions. The free parameters of this model are
defined as $\Delta = (26 \pm 2)$ meV and $v_{F} = (3.0 \pm 0.2)
\times10^{5}$ m/s, which can be considered as averaged values over
the entire in-plane part of the nodal line. Both values are well
consistent with the ones obtained by other methods and reported in
literature. The fact that such a simple model accounts for the
complete magneto-optical spectrum indicates that $\Delta$ and
$v_{F}$ do not vary appreciably along (the in-plane portion of) the
nodal line. This also evidences that other bands do not contribute
to the corresponding ac transport. Possible origins of the third
observed mode are discussed and more theory output is called for.

\begin{acknowledgments}
The authors acknowledge fruitful discussions with Chao Zhang
(Wollongong), Milan Orlita (Grenoble), and Sascha Polatkan
(Stuttgart) and the technical support by Gabriele Untereiner
(Stuttgart). The work was supported by the Deutsche
Forschungsgesellschaft (DFG) via DR228/51-1 and the Max Planck
Society. E.U. acknowledges the support by the ESF and by the
Ministry of Science, Research, and Arts of Baden-W\"{u}rttemberg.
Work at Princeton was supported by NSF through the Princeton Center
for Complex Materials, a Materials Research Science and Engineering
Center (grant No. DMR-1420541).
\end{acknowledgments}

\bibliographystyle{apsrev4-1}
\bibliography{ZrSiS_references}

\begin{thebibliography}{27}%
\makeatletter
\providecommand \@ifxundefined [1]{%
 \@ifx{#1\undefined}
}%
\providecommand \@ifnum [1]{%
 \ifnum #1\expandafter \@firstoftwo
 \else \expandafter \@secondoftwo
 \fi
}%
\providecommand \@ifx [1]{%
 \ifx #1\expandafter \@firstoftwo
 \else \expandafter \@secondoftwo
 \fi
}%
\providecommand \natexlab [1]{#1}%
\providecommand \enquote  [1]{``#1''}%
\providecommand \bibnamefont  [1]{#1}%
\providecommand \bibfnamefont [1]{#1}%
\providecommand \citenamefont [1]{#1}%
\providecommand \href@noop [0]{\@secondoftwo}%
\providecommand \href [0]{\begingroup \@sanitize@url \@href}%
\providecommand \@href[1]{\@@startlink{#1}\@@href}%
\providecommand \@@href[1]{\endgroup#1\@@endlink}%
\providecommand \@sanitize@url [0]{\catcode `\\12\catcode `\$12\catcode
  `\&12\catcode `\#12\catcode `\^12\catcode `\_12\catcode `\%12\relax}%
\providecommand \@@startlink[1]{}%
\providecommand \@@endlink[0]{}%
\providecommand \url  [0]{\begingroup\@sanitize@url \@url }%
\providecommand \@url [1]{\endgroup\@href {#1}{\urlprefix }}%
\providecommand \urlprefix  [0]{URL }%
\providecommand \Eprint [0]{\href }%
\providecommand \doibase [0]{http://dx.doi.org/}%
\providecommand \selectlanguage [0]{\@gobble}%
\providecommand \bibinfo  [0]{\@secondoftwo}%
\providecommand \bibfield  [0]{\@secondoftwo}%
\providecommand \translation [1]{[#1]}%
\providecommand \BibitemOpen [0]{}%
\providecommand \bibitemStop [0]{}%
\providecommand \bibitemNoStop [0]{.\EOS\space}%
\providecommand \EOS [0]{\spacefactor3000\relax}%
\providecommand \BibitemShut  [1]{\csname bibitem#1\endcsname}%
\let\auto@bib@innerbib\@empty
%</preamble>
\bibitem [{\citenamefont {Burkov}\ \emph {et~al.}(2011)\citenamefont {Burkov},
  \citenamefont {Hook},\ and\ \citenamefont {Balents}}]{Burkov2011}%
  \BibitemOpen
  \bibfield  {author} {\bibinfo {author} {\bibfnamefont {A.~A.}\ \bibnamefont
  {Burkov}}, \bibinfo {author} {\bibfnamefont {M.~D.}\ \bibnamefont {Hook}}, \
  and\ \bibinfo {author} {\bibfnamefont {L.}~\bibnamefont {Balents}},\ }\href
  {\doibase 10.1103/PhysRevB.84.235126} {\bibfield  {journal} {\bibinfo
  {journal} {Phys. Rev. B}\ }\textbf {\bibinfo {volume} {84}},\ \bibinfo
  {pages} {235126} (\bibinfo {year} {2011})}\BibitemShut {NoStop}%
\bibitem [{\citenamefont {Fang}\ \emph {et~al.}(2016)\citenamefont {Fang},
  \citenamefont {Weng}, \citenamefont {Dai},\ and\ \citenamefont
  {Fang}}]{Fang2014}%
  \BibitemOpen
  \bibfield  {author} {\bibinfo {author} {\bibfnamefont {C.}~\bibnamefont
  {Fang}}, \bibinfo {author} {\bibfnamefont {H.}~\bibnamefont {Weng}}, \bibinfo
  {author} {\bibfnamefont {X.}~\bibnamefont {Dai}}, \ and\ \bibinfo {author}
  {\bibfnamefont {Z.}~\bibnamefont {Fang}},\ }\href {\doibase
  10.1088/1674-1056/25/11/117106} {\bibfield  {journal} {\bibinfo  {journal}
  {Chin. Phys. B}\ }\textbf {\bibinfo {volume} {25}},\ \bibinfo {eid} {117106}
  (\bibinfo {year} {2016})}\BibitemShut {NoStop}%
\bibitem [{\citenamefont {Bian}\ \emph {et~al.}(2016)\citenamefont {Bian},
  \citenamefont {Chang}, \citenamefont {Sankar}, \citenamefont {Xu},
  \citenamefont {Zheng}, \citenamefont {Neupert}, \citenamefont {Chiu},
  \citenamefont {Huang}, \citenamefont {Chang}, \citenamefont {Belopolski},
  \citenamefont {Sanchez}, \citenamefont {Neupane}, \citenamefont {Alidoust},
  \citenamefont {Liu}, \citenamefont {Wang}, \citenamefont {Lee}, \citenamefont
  {Jeng}, \citenamefont {Zhang}, \citenamefont {Yuan}, \citenamefont {Jia},
  \citenamefont {Bansil}, \citenamefont {Chou}, \citenamefont {Lin},\ and\
  \citenamefont {Hasan}}]{Bian2016}%
  \BibitemOpen
  \bibfield  {author} {\bibinfo {author} {\bibfnamefont {G.}~\bibnamefont
  {Bian}}, \bibinfo {author} {\bibfnamefont {T.-R.}\ \bibnamefont {Chang}},
  \bibinfo {author} {\bibfnamefont {R.}~\bibnamefont {Sankar}}, \bibinfo
  {author} {\bibfnamefont {S.-Y.}\ \bibnamefont {Xu}}, \bibinfo {author}
  {\bibfnamefont {H.}~\bibnamefont {Zheng}}, \bibinfo {author} {\bibfnamefont
  {T.}~\bibnamefont {Neupert}}, \bibinfo {author} {\bibfnamefont {C.-K.}\
  \bibnamefont {Chiu}}, \bibinfo {author} {\bibfnamefont {S.-M.}\ \bibnamefont
  {Huang}}, \bibinfo {author} {\bibfnamefont {G.}~\bibnamefont {Chang}},
  \bibinfo {author} {\bibfnamefont {I.}~\bibnamefont {Belopolski}}, \bibinfo
  {author} {\bibfnamefont {D.~S.}\ \bibnamefont {Sanchez}}, \bibinfo {author}
  {\bibfnamefont {M.}~\bibnamefont {Neupane}}, \bibinfo {author} {\bibfnamefont
  {N.}~\bibnamefont {Alidoust}}, \bibinfo {author} {\bibfnamefont
  {C.}~\bibnamefont {Liu}}, \bibinfo {author} {\bibfnamefont {B.}~\bibnamefont
  {Wang}}, \bibinfo {author} {\bibfnamefont {C.-C.}\ \bibnamefont {Lee}},
  \bibinfo {author} {\bibfnamefont {H.-T.}\ \bibnamefont {Jeng}}, \bibinfo
  {author} {\bibfnamefont {C.}~\bibnamefont {Zhang}}, \bibinfo {author}
  {\bibfnamefont {Z.}~\bibnamefont {Yuan}}, \bibinfo {author} {\bibfnamefont
  {S.}~\bibnamefont {Jia}}, \bibinfo {author} {\bibfnamefont {A.}~\bibnamefont
  {Bansil}}, \bibinfo {author} {\bibfnamefont {F.}~\bibnamefont {Chou}},
  \bibinfo {author} {\bibfnamefont {H.}~\bibnamefont {Lin}}, \ and\ \bibinfo
  {author} {\bibfnamefont {M.~Z.}\ \bibnamefont {Hasan}},\ }\href
  {https://doi.org/10.1038/ncomms10556} {\bibfield  {journal} {\bibinfo
  {journal} {Nat. Comm.}\ }\textbf {\bibinfo {volume} {7}},\ \bibinfo {pages}
  {10556} (\bibinfo {year} {2016})}\BibitemShut {NoStop}%
\bibitem [{\citenamefont {Yu}\ \emph {et~al.}(2015)\citenamefont {Yu},
  \citenamefont {Weng}, \citenamefont {Fang}, \citenamefont {Dai},\ and\
  \citenamefont {Hu}}]{Yu2015}%
  \BibitemOpen
  \bibfield  {author} {\bibinfo {author} {\bibfnamefont {R.}~\bibnamefont
  {Yu}}, \bibinfo {author} {\bibfnamefont {H.}~\bibnamefont {Weng}}, \bibinfo
  {author} {\bibfnamefont {Z.}~\bibnamefont {Fang}}, \bibinfo {author}
  {\bibfnamefont {X.}~\bibnamefont {Dai}}, \ and\ \bibinfo {author}
  {\bibfnamefont {X.}~\bibnamefont {Hu}},\ }\href {\doibase
  10.1103/PhysRevLett.115.036807} {\bibfield  {journal} {\bibinfo  {journal}
  {Phys. Rev. Lett.}\ }\textbf {\bibinfo {volume} {115}},\ \bibinfo {pages}
  {036807} (\bibinfo {year} {2015})}\BibitemShut {NoStop}%
\bibitem [{\citenamefont {Chen}\ \emph {et~al.}(2015)\citenamefont {Chen},
  \citenamefont {Lu},\ and\ \citenamefont {Kee}}]{Chen2015}%
  \BibitemOpen
  \bibfield  {author} {\bibinfo {author} {\bibfnamefont {Y.}~\bibnamefont
  {Chen}}, \bibinfo {author} {\bibfnamefont {Y.-M.}\ \bibnamefont {Lu}}, \ and\
  \bibinfo {author} {\bibfnamefont {H.-Y.}\ \bibnamefont {Kee}},\ }\href
  {https://doi.org/10.1038/ncomms7593} {\bibfield  {journal} {\bibinfo
  {journal} {Nat. Comm.}\ }\textbf {\bibinfo {volume} {6}},\ \bibinfo {pages}
  {6593} (\bibinfo {year} {2015})}\BibitemShut {NoStop}%
\bibitem [{\citenamefont {Xie}\ \emph {et~al.}(2015)\citenamefont {Xie},
  \citenamefont {Schoop}, \citenamefont {Seibel}, \citenamefont {Gibson},
  \citenamefont {Xie},\ and\ \citenamefont {Cava}}]{Xie2015}%
  \BibitemOpen
  \bibfield  {author} {\bibinfo {author} {\bibfnamefont {L.~S.}\ \bibnamefont
  {Xie}}, \bibinfo {author} {\bibfnamefont {L.~M.}\ \bibnamefont {Schoop}},
  \bibinfo {author} {\bibfnamefont {E.~M.}\ \bibnamefont {Seibel}}, \bibinfo
  {author} {\bibfnamefont {Q.~D.}\ \bibnamefont {Gibson}}, \bibinfo {author}
  {\bibfnamefont {W.}~\bibnamefont {Xie}}, \ and\ \bibinfo {author}
  {\bibfnamefont {R.~J.}\ \bibnamefont {Cava}},\ }\href
  {https://doi.org/10.1063/1.4926545} {\bibfield  {journal} {\bibinfo
  {journal} {APL Mater.}\ }\textbf {\bibinfo {volume} {3}},\ \bibinfo {pages}
  {083602} (\bibinfo {year} {2015})}\BibitemShut {NoStop}%
\bibitem [{\citenamefont {Schoop}\ \emph {et~al.}(2016)\citenamefont {Schoop},
  \citenamefont {Ali}, \citenamefont {Straßer}, \citenamefont {Topp},
  \citenamefont {Varykhalov}, \citenamefont {Marchenko}, \citenamefont
  {Duppel}, \citenamefont {Parkin}, \citenamefont {Lotsch},\ and\ \citenamefont
  {Ast}}]{Schoop2016}%
  \BibitemOpen
  \bibfield  {author} {\bibinfo {author} {\bibfnamefont {L.~M.}\ \bibnamefont
  {Schoop}}, \bibinfo {author} {\bibfnamefont {M.~N.}\ \bibnamefont {Ali}},
  \bibinfo {author} {\bibfnamefont {C.}~\bibnamefont {Straßer}}, \bibinfo
  {author} {\bibfnamefont {A.}~\bibnamefont {Topp}}, \bibinfo {author}
  {\bibfnamefont {A.}~\bibnamefont {Varykhalov}}, \bibinfo {author}
  {\bibfnamefont {D.}~\bibnamefont {Marchenko}}, \bibinfo {author}
  {\bibfnamefont {V.}~\bibnamefont {Duppel}}, \bibinfo {author} {\bibfnamefont
  {S.~S.~P.}\ \bibnamefont {Parkin}}, \bibinfo {author} {\bibfnamefont {B.~V.}\
  \bibnamefont {Lotsch}}, \ and\ \bibinfo {author} {\bibfnamefont {C.~R.}\
  \bibnamefont {Ast}},\ }\href {https://doi.org/10.1038/ncomms11696} {\bibfield
   {journal} {\bibinfo  {journal} {Nat. Comm.}\ }\textbf {\bibinfo {volume}
  {7}},\ \bibinfo {pages} {11696} (\bibinfo {year} {2016})}\BibitemShut
  {NoStop}%
\bibitem [{\citenamefont {Chen}\ \emph
  {et~al.}(2017{\natexlab{a}})\citenamefont {Chen}, \citenamefont {Xu},
  \citenamefont {Jiang}, \citenamefont {Wu}, \citenamefont {Qi}, \citenamefont
  {Yang}, \citenamefont {Wang}, \citenamefont {Sun}, \citenamefont
  {Schr\"oter}, \citenamefont {Yang}, \citenamefont {Schoop}, \citenamefont
  {Lv}, \citenamefont {Zhou}, \citenamefont {Chen}, \citenamefont {Yao},
  \citenamefont {Lu}, \citenamefont {Chen}, \citenamefont {Felser},
  \citenamefont {Yan}, \citenamefont {Liu},\ and\ \citenamefont
  {Chen}}]{Chen2017}%
  \BibitemOpen
  \bibfield  {author} {\bibinfo {author} {\bibfnamefont {C.}~\bibnamefont
  {Chen}}, \bibinfo {author} {\bibfnamefont {X.}~\bibnamefont {Xu}}, \bibinfo
  {author} {\bibfnamefont {J.}~\bibnamefont {Jiang}}, \bibinfo {author}
  {\bibfnamefont {S.-C.}\ \bibnamefont {Wu}}, \bibinfo {author} {\bibfnamefont
  {Y.~P.}\ \bibnamefont {Qi}}, \bibinfo {author} {\bibfnamefont {L.~X.}\
  \bibnamefont {Yang}}, \bibinfo {author} {\bibfnamefont {M.~X.}\ \bibnamefont
  {Wang}}, \bibinfo {author} {\bibfnamefont {Y.}~\bibnamefont {Sun}}, \bibinfo
  {author} {\bibfnamefont {N.~B.~M.}\ \bibnamefont {Schr\"oter}}, \bibinfo
  {author} {\bibfnamefont {H.~F.}\ \bibnamefont {Yang}}, \bibinfo {author}
  {\bibfnamefont {L.~M.}\ \bibnamefont {Schoop}}, \bibinfo {author}
  {\bibfnamefont {Y.~Y.}\ \bibnamefont {Lv}}, \bibinfo {author} {\bibfnamefont
  {J.}~\bibnamefont {Zhou}}, \bibinfo {author} {\bibfnamefont {Y.~B.}\
  \bibnamefont {Chen}}, \bibinfo {author} {\bibfnamefont {S.~H.}\ \bibnamefont
  {Yao}}, \bibinfo {author} {\bibfnamefont {M.~H.}\ \bibnamefont {Lu}},
  \bibinfo {author} {\bibfnamefont {Y.~F.}\ \bibnamefont {Chen}}, \bibinfo
  {author} {\bibfnamefont {C.}~\bibnamefont {Felser}}, \bibinfo {author}
  {\bibfnamefont {B.~H.}\ \bibnamefont {Yan}}, \bibinfo {author} {\bibfnamefont
  {Z.~K.}\ \bibnamefont {Liu}}, \ and\ \bibinfo {author} {\bibfnamefont
  {Y.~L.}\ \bibnamefont {Chen}},\ }\href
  {https://link.aps.org/doi/10.1103/PhysRevB.95.125126} {\bibfield  {journal}
  {\bibinfo  {journal} {Phys. Rev. B}\ }\textbf {\bibinfo {volume} {95}},\
  \bibinfo {pages} {125126} (\bibinfo {year} {2017}{\natexlab{a}})}\BibitemShut
  {NoStop}%
\bibitem [{\citenamefont {Neupane}\ \emph {et~al.}(2016)\citenamefont
  {Neupane}, \citenamefont {Belopolski}, \citenamefont {Hosen}, \citenamefont
  {Sanchez}, \citenamefont {Sankar}, \citenamefont {Szlawska}, \citenamefont
  {Xu}, \citenamefont {Dimitri}, \citenamefont {Dhakal}, \citenamefont
  {Maldonado}, \citenamefont {Oppeneer}, \citenamefont {Kaczorowski},
  \citenamefont {Chou}, \citenamefont {Hasan},\ and\ \citenamefont
  {Durakiewicz}}]{Neupane2016}%
  \BibitemOpen
  \bibfield  {author} {\bibinfo {author} {\bibfnamefont {M.}~\bibnamefont
  {Neupane}}, \bibinfo {author} {\bibfnamefont {I.}~\bibnamefont {Belopolski}},
  \bibinfo {author} {\bibfnamefont {M.~M.}\ \bibnamefont {Hosen}}, \bibinfo
  {author} {\bibfnamefont {D.~S.}\ \bibnamefont {Sanchez}}, \bibinfo {author}
  {\bibfnamefont {R.}~\bibnamefont {Sankar}}, \bibinfo {author} {\bibfnamefont
  {M.}~\bibnamefont {Szlawska}}, \bibinfo {author} {\bibfnamefont {S.-Y.}\
  \bibnamefont {Xu}}, \bibinfo {author} {\bibfnamefont {K.}~\bibnamefont
  {Dimitri}}, \bibinfo {author} {\bibfnamefont {N.}~\bibnamefont {Dhakal}},
  \bibinfo {author} {\bibfnamefont {P.}~\bibnamefont {Maldonado}}, \bibinfo
  {author} {\bibfnamefont {P.~M.}\ \bibnamefont {Oppeneer}}, \bibinfo {author}
  {\bibfnamefont {D.}~\bibnamefont {Kaczorowski}}, \bibinfo {author}
  {\bibfnamefont {F.}~\bibnamefont {Chou}}, \bibinfo {author} {\bibfnamefont
  {M.~Z.}\ \bibnamefont {Hasan}}, \ and\ \bibinfo {author} {\bibfnamefont
  {T.}~\bibnamefont {Durakiewicz}},\ }\href {\doibase
  10.1103/PhysRevB.93.201104} {\bibfield  {journal} {\bibinfo  {journal} {Phys.
  Rev. B}\ }\textbf {\bibinfo {volume} {93}},\ \bibinfo {pages} {201104}
  (\bibinfo {year} {2016})}\BibitemShut {NoStop}%
\bibitem [{\citenamefont {Ali}\ \emph {et~al.}(2016)\citenamefont {Ali},
  \citenamefont {Schoop}, \citenamefont {Garg}, \citenamefont {Lippmann},
  \citenamefont {Lara}, \citenamefont {Lotsch},\ and\ \citenamefont
  {Parkin}}]{Ali2016}%
  \BibitemOpen
  \bibfield  {author} {\bibinfo {author} {\bibfnamefont {M.~N.}\ \bibnamefont
  {Ali}}, \bibinfo {author} {\bibfnamefont {L.~M.}\ \bibnamefont {Schoop}},
  \bibinfo {author} {\bibfnamefont {C.}~\bibnamefont {Garg}}, \bibinfo {author}
  {\bibfnamefont {J.~M.}\ \bibnamefont {Lippmann}}, \bibinfo {author}
  {\bibfnamefont {E.}~\bibnamefont {Lara}}, \bibinfo {author} {\bibfnamefont
  {B.}~\bibnamefont {Lotsch}}, \ and\ \bibinfo {author} {\bibfnamefont
  {S.~S.~P.}\ \bibnamefont {Parkin}},\ }\href
  {https://advances.sciencemag.org/content/2/12/e1601742} {\bibfield  {journal}
  {\bibinfo  {journal} {Sci. Adv.}\ }\textbf {\bibinfo {volume} {2}} (\bibinfo
  {year} {2016})}\BibitemShut {NoStop}%
\bibitem [{\citenamefont {Pezzini}\ \emph {et~al.}(2017)\citenamefont
  {Pezzini}, \citenamefont {van Delft}, \citenamefont {Schoop}, \citenamefont
  {Lotsch}, \citenamefont {Carrington}, \citenamefont {Katsnelson},
  \citenamefont {Hussey},\ and\ \citenamefont {Wiedmann}}]{Pezzini2017}%
  \BibitemOpen
  \bibfield  {author} {\bibinfo {author} {\bibfnamefont {S.}~\bibnamefont
  {Pezzini}}, \bibinfo {author} {\bibfnamefont {M.~R.}\ \bibnamefont {van
  Delft}}, \bibinfo {author} {\bibfnamefont {L.~M.}\ \bibnamefont {Schoop}},
  \bibinfo {author} {\bibfnamefont {B.~V.}\ \bibnamefont {Lotsch}}, \bibinfo
  {author} {\bibfnamefont {A.}~\bibnamefont {Carrington}}, \bibinfo {author}
  {\bibfnamefont {M.~I.}\ \bibnamefont {Katsnelson}}, \bibinfo {author}
  {\bibfnamefont {N.}~\bibnamefont {Hussey}}, \ and\ \bibinfo {author}
  {\bibfnamefont {S.}~\bibnamefont {Wiedmann}},\ }\href
  {https://doi.org/10.1038/nphys4306} {\bibfield  {journal} {\bibinfo
  {journal} {Nat. Phys.}\ }\textbf {\bibinfo {volume} {14}},\ \bibinfo {pages}
  {178} (\bibinfo {year} {2017})}\BibitemShut {NoStop}%
\bibitem [{\citenamefont {Matusiak}\ \emph {et~al.}(2017)\citenamefont
  {Matusiak}, \citenamefont {Cooper},\ and\ \citenamefont
  {Kaczorowski}}]{Matusiak2017}%
  \BibitemOpen
  \bibfield  {author} {\bibinfo {author} {\bibfnamefont {M.}~\bibnamefont
  {Matusiak}}, \bibinfo {author} {\bibfnamefont {J.~R.}\ \bibnamefont
  {Cooper}}, \ and\ \bibinfo {author} {\bibfnamefont {D.}~\bibnamefont
  {Kaczorowski}},\ }\href {https://doi.org/10.1038/ncomms15219} {\bibfield
  {journal} {\bibinfo  {journal} {Nat. Comm.}\ }\textbf {\bibinfo {volume}
  {8}},\ \bibinfo {pages} {15219} (\bibinfo {year} {2017})}\BibitemShut
  {NoStop}%
\bibitem [{\citenamefont {Hu}\ \emph {et~al.}(2017)\citenamefont {Hu},
  \citenamefont {Tang}, \citenamefont {Liu}, \citenamefont {Zhu}, \citenamefont
  {Wei},\ and\ \citenamefont {Mao}}]{Hu2017}%
  \BibitemOpen
  \bibfield  {author} {\bibinfo {author} {\bibfnamefont {J.}~\bibnamefont
  {Hu}}, \bibinfo {author} {\bibfnamefont {Z.}~\bibnamefont {Tang}}, \bibinfo
  {author} {\bibfnamefont {J.}~\bibnamefont {Liu}}, \bibinfo {author}
  {\bibfnamefont {Y.}~\bibnamefont {Zhu}}, \bibinfo {author} {\bibfnamefont
  {J.}~\bibnamefont {Wei}}, \ and\ \bibinfo {author} {\bibfnamefont
  {Z.}~\bibnamefont {Mao}},\ }\href {\doibase 10.1103/PhysRevB.96.045127}
  {\bibfield  {journal} {\bibinfo  {journal} {Phys. Rev. B}\ }\textbf {\bibinfo
  {volume} {96}},\ \bibinfo {pages} {045127} (\bibinfo {year}
  {2017})}\BibitemShut {NoStop}%
\bibitem [{\citenamefont {Schilling}\ \emph {et~al.}(2017)\citenamefont
  {Schilling}, \citenamefont {Schoop}, \citenamefont {Lotsch}, \citenamefont
  {Dressel},\ and\ \citenamefont {Pronin}}]{Schilling2017}%
  \BibitemOpen
  \bibfield  {author} {\bibinfo {author} {\bibfnamefont {M.~B.}\ \bibnamefont
  {Schilling}}, \bibinfo {author} {\bibfnamefont {L.~M.}\ \bibnamefont
  {Schoop}}, \bibinfo {author} {\bibfnamefont {B.~V.}\ \bibnamefont {Lotsch}},
  \bibinfo {author} {\bibfnamefont {M.}~\bibnamefont {Dressel}}, \ and\
  \bibinfo {author} {\bibfnamefont {A.~V.}\ \bibnamefont {Pronin}},\ }\href
  {\doibase 10.1103/PhysRevLett.119.187401} {\bibfield  {journal} {\bibinfo
  {journal} {Phys. Rev. Lett.}\ }\textbf {\bibinfo {volume} {119}},\ \bibinfo
  {pages} {187401} (\bibinfo {year} {2017})}\BibitemShut {NoStop}%
\bibitem [{\citenamefont {Orlita}\ \emph {et~al.}(2011)\citenamefont {Orlita},
  \citenamefont {Faugeras}, \citenamefont {Grill}, \citenamefont {Wysmolek},
  \citenamefont {Strupinski}, \citenamefont {Berger}, \citenamefont {de~Heer},
  \citenamefont {Martinez},\ and\ \citenamefont {Potemski}}]{Orlita2011}%
  \BibitemOpen
  \bibfield  {author} {\bibinfo {author} {\bibfnamefont {M.}~\bibnamefont
  {Orlita}}, \bibinfo {author} {\bibfnamefont {C.}~\bibnamefont {Faugeras}},
  \bibinfo {author} {\bibfnamefont {R.}~\bibnamefont {Grill}}, \bibinfo
  {author} {\bibfnamefont {A.}~\bibnamefont {Wysmolek}}, \bibinfo {author}
  {\bibfnamefont {W.}~\bibnamefont {Strupinski}}, \bibinfo {author}
  {\bibfnamefont {C.}~\bibnamefont {Berger}}, \bibinfo {author} {\bibfnamefont
  {W.~A.}\ \bibnamefont {de~Heer}}, \bibinfo {author} {\bibfnamefont
  {G.}~\bibnamefont {Martinez}}, \ and\ \bibinfo {author} {\bibfnamefont
  {M.}~\bibnamefont {Potemski}},\ }\href {\doibase
  10.1103/PhysRevLett.107.216603} {\bibfield  {journal} {\bibinfo  {journal}
  {Phys. Rev. Lett.}\ }\textbf {\bibinfo {volume} {107}},\ \bibinfo {pages}
  {216603} (\bibinfo {year} {2011})}\BibitemShut {NoStop}%
\bibitem [{\citenamefont {Orlita}\ \emph {et~al.}(2012)\citenamefont {Orlita},
  \citenamefont {Crassee}, \citenamefont {Faugeras}, \citenamefont {Kuzmenko},
  \citenamefont {Fromm}, \citenamefont {Ostler}, \citenamefont {Seyller},
  \citenamefont {G.}, \citenamefont {M.},\ and\ \citenamefont
  {M.}}]{Orlita2012}%
  \BibitemOpen
  \bibfield  {author} {\bibinfo {author} {\bibfnamefont {M.}~\bibnamefont
  {Orlita}}, \bibinfo {author} {\bibfnamefont {I.}~\bibnamefont {Crassee}},
  \bibinfo {author} {\bibfnamefont {C.}~\bibnamefont {Faugeras}}, \bibinfo
  {author} {\bibfnamefont {A.~B.}\ \bibnamefont {Kuzmenko}}, \bibinfo {author}
  {\bibfnamefont {F.}~\bibnamefont {Fromm}}, \bibinfo {author} {\bibfnamefont
  {M.}~\bibnamefont {Ostler}}, \bibinfo {author} {\bibfnamefont
  {T.}~\bibnamefont {Seyller}}, \bibinfo {author} {\bibfnamefont
  {M.}~\bibnamefont {G.}}, \bibinfo {author} {\bibfnamefont {P.}~\bibnamefont
  {M.}}, \ and\ \bibinfo {author} {\bibfnamefont {P.}~\bibnamefont {M.}},\
  }\href {\doibase 10.1088/1367-2630/14/9/095008} {\bibfield  {journal}
  {\bibinfo  {journal} {New. J. Phys.}\ }\textbf {\bibinfo {volume} {14}},\
  \bibinfo {pages} {095008} (\bibinfo {year} {2012})}\BibitemShut {NoStop}%
\bibitem [{\citenamefont {Chen}\ \emph
  {et~al.}(2017{\natexlab{b}})\citenamefont {Chen}, \citenamefont {Chen},
  \citenamefont {Zhong}, \citenamefont {Schneeloch}, \citenamefont {Zhang},
  \citenamefont {Huang}, \citenamefont {Qu}, \citenamefont {Yu}, \citenamefont
  {Li}, \citenamefont {Gu},\ and\ \citenamefont {Wang}}]{Chen2017a}%
  \BibitemOpen
  \bibfield  {author} {\bibinfo {author} {\bibfnamefont {Z.-G.}\ \bibnamefont
  {Chen}}, \bibinfo {author} {\bibfnamefont {R.~Y.}\ \bibnamefont {Chen}},
  \bibinfo {author} {\bibfnamefont {R.~D.}\ \bibnamefont {Zhong}}, \bibinfo
  {author} {\bibfnamefont {J.}~\bibnamefont {Schneeloch}}, \bibinfo {author}
  {\bibfnamefont {C.}~\bibnamefont {Zhang}}, \bibinfo {author} {\bibfnamefont
  {Y.}~\bibnamefont {Huang}}, \bibinfo {author} {\bibfnamefont
  {F.}~\bibnamefont {Qu}}, \bibinfo {author} {\bibfnamefont {R.}~\bibnamefont
  {Yu}}, \bibinfo {author} {\bibfnamefont {Q.}~\bibnamefont {Li}}, \bibinfo
  {author} {\bibfnamefont {G.~D.}\ \bibnamefont {Gu}}, \ and\ \bibinfo {author}
  {\bibfnamefont {N.~L.}\ \bibnamefont {Wang}},\ }\href
  {https://www.pnas.org/content/114/5/816} {\bibfield  {journal} {\bibinfo
  {journal} {Proc. Natl. Acad. Sci.}\ }\textbf {\bibinfo {volume} {114}},\
  \bibinfo {pages} {816} (\bibinfo {year} {2017}{\natexlab{b}})}\BibitemShut
  {NoStop}%
\bibitem [{\citenamefont {Chen}\ \emph
  {et~al.}(2014{\natexlab{a}})\citenamefont {Chen}, \citenamefont {Shi},
  \citenamefont {Yang}, \citenamefont {Lu}, \citenamefont {Lai}, \citenamefont
  {Yan}, \citenamefont {Wang}, \citenamefont {Zhang},\ and\ \citenamefont
  {Li}}]{Chen2014a}%
  \BibitemOpen
  \bibfield  {author} {\bibinfo {author} {\bibfnamefont {Z.-G.}\ \bibnamefont
  {Chen}}, \bibinfo {author} {\bibfnamefont {Z.}~\bibnamefont {Shi}}, \bibinfo
  {author} {\bibfnamefont {W.}~\bibnamefont {Yang}}, \bibinfo {author}
  {\bibfnamefont {X.}~\bibnamefont {Lu}}, \bibinfo {author} {\bibfnamefont
  {Y.}~\bibnamefont {Lai}}, \bibinfo {author} {\bibfnamefont {H.}~\bibnamefont
  {Yan}}, \bibinfo {author} {\bibfnamefont {F.}~\bibnamefont {Wang}}, \bibinfo
  {author} {\bibfnamefont {G.}~\bibnamefont {Zhang}}, \ and\ \bibinfo {author}
  {\bibfnamefont {Z.}~\bibnamefont {Li}},\ }\href
  {https://doi.org/10.1038/ncomms5461} {\bibfield  {journal} {\bibinfo
  {journal} {Nat. Comm.}\ }\textbf {\bibinfo {volume} {5}},\ \bibinfo {pages}
  {4461} (\bibinfo {year} {2014}{\natexlab{a}})}\BibitemShut {NoStop}%
\bibitem [{\citenamefont {Shao}\ \emph {et~al.}(2019)\citenamefont {Shao},
  \citenamefont {Sun}, \citenamefont {Wang}, \citenamefont {Xu}, \citenamefont
  {Sankar}, \citenamefont {Breindel}, \citenamefont {Cao}, \citenamefont
  {Fogler}, \citenamefont {Millis}, \citenamefont {Chou}, \citenamefont {Li},
  \citenamefont {Timusk}, \citenamefont {Maple},\ and\ \citenamefont
  {Basov}}]{Shao2019}%
  \BibitemOpen
  \bibfield  {author} {\bibinfo {author} {\bibfnamefont {Y.}~\bibnamefont
  {Shao}}, \bibinfo {author} {\bibfnamefont {Z.}~\bibnamefont {Sun}}, \bibinfo
  {author} {\bibfnamefont {Y.}~\bibnamefont {Wang}}, \bibinfo {author}
  {\bibfnamefont {C.}~\bibnamefont {Xu}}, \bibinfo {author} {\bibfnamefont
  {R.}~\bibnamefont {Sankar}}, \bibinfo {author} {\bibfnamefont {A.~J.}\
  \bibnamefont {Breindel}}, \bibinfo {author} {\bibfnamefont {C.}~\bibnamefont
  {Cao}}, \bibinfo {author} {\bibfnamefont {M.~M.}\ \bibnamefont {Fogler}},
  \bibinfo {author} {\bibfnamefont {A.~J.}\ \bibnamefont {Millis}}, \bibinfo
  {author} {\bibfnamefont {F.}~\bibnamefont {Chou}}, \bibinfo {author}
  {\bibfnamefont {Z.}~\bibnamefont {Li}}, \bibinfo {author} {\bibfnamefont
  {T.}~\bibnamefont {Timusk}}, \bibinfo {author} {\bibfnamefont {M.~B.}\
  \bibnamefont {Maple}}, \ and\ \bibinfo {author} {\bibfnamefont {D.~N.}\
  \bibnamefont {Basov}},\ }\href {https://www.pnas.org/content/116/4/1168}
  {\bibfield  {journal} {\bibinfo  {journal} {Proc. Natl. Acad. Sci.}\ }\textbf
  {\bibinfo {volume} {116}},\ \bibinfo {pages} {1168} (\bibinfo {year}
  {2019})}\BibitemShut {NoStop}%
\bibitem [{\citenamefont {Palik}\ and\ \citenamefont
  {Furdyna}(1970)}]{Palik1970}%
  \BibitemOpen
  \bibfield  {author} {\bibinfo {author} {\bibfnamefont {E.~D.}\ \bibnamefont
  {Palik}}\ and\ \bibinfo {author} {\bibfnamefont {J.~K.}\ \bibnamefont
  {Furdyna}},\ }\href {\doibase 10.1088/0034-4885/33/3/307} {\bibfield
  {journal} {\bibinfo  {journal} {Rep. Prog. Phys.}\ }\textbf {\bibinfo
  {volume} {33}},\ \bibinfo {pages} {1193} (\bibinfo {year}
  {1970})}\BibitemShut {NoStop}%
\bibitem [{\citenamefont {Tanner}(2015)}]{Tanner2015}%
  \BibitemOpen
  \bibfield  {author} {\bibinfo {author} {\bibfnamefont {D.~B.}\ \bibnamefont
  {Tanner}},\ }\href {\doibase 10.1103/PhysRevB.91.035123} {\bibfield
  {journal} {\bibinfo  {journal} {Phys. Rev. B}\ }\textbf {\bibinfo {volume}
  {91}},\ \bibinfo {pages} {035123} (\bibinfo {year} {2015})}\BibitemShut
  {NoStop}%
\bibitem [{\citenamefont {Chen}\ \emph
  {et~al.}(2014{\natexlab{b}})\citenamefont {Chen}, \citenamefont {Wallbank},
  \citenamefont {Patel}, \citenamefont {Mucha-Kruczy\ifmmode~\acute{n}\else
  \'{n}\fi{}ski}, \citenamefont {McCann},\ and\ \citenamefont
  {Fal'ko}}]{Chen2014b}%
  \BibitemOpen
  \bibfield  {author} {\bibinfo {author} {\bibfnamefont {X.}~\bibnamefont
  {Chen}}, \bibinfo {author} {\bibfnamefont {J.~R.}\ \bibnamefont {Wallbank}},
  \bibinfo {author} {\bibfnamefont {A.~A.}\ \bibnamefont {Patel}}, \bibinfo
  {author} {\bibfnamefont {M.}~\bibnamefont
  {Mucha-Kruczy\ifmmode~\acute{n}\else \'{n}\fi{}ski}}, \bibinfo {author}
  {\bibfnamefont {E.}~\bibnamefont {McCann}}, \ and\ \bibinfo {author}
  {\bibfnamefont {V.~I.}\ \bibnamefont {Fal'ko}},\ }\href {\doibase
  10.1103/PhysRevB.89.075401} {\bibfield  {journal} {\bibinfo  {journal} {Phys.
  Rev. B}\ }\textbf {\bibinfo {volume} {89}},\ \bibinfo {pages} {075401}
  (\bibinfo {year} {2014}{\natexlab{b}})}\BibitemShut {NoStop}%
\bibitem [{\citenamefont {Dressel}\ and\ \citenamefont
  {Gr{\"{u}}ner}(2002)}]{Dressel2002}%
  \BibitemOpen
  \bibfield  {author} {\bibinfo {author} {\bibfnamefont {M.}~\bibnamefont
  {Dressel}}\ and\ \bibinfo {author} {\bibfnamefont {G.}~\bibnamefont
  {Gr{\"{u}}ner}},\ }\href@noop {} {\emph {\bibinfo {title} {Electrodynamics of
  Solids}}}\ (\bibinfo  {publisher} {Cambridge University Press, Cambridge},\
  \bibinfo {year} {2002})\BibitemShut {NoStop}%
\bibitem [{\citenamefont {Lodge}\ \emph {et~al.}(2017)\citenamefont {Lodge},
  \citenamefont {Chang}, \citenamefont {Huang}, \citenamefont {Singh},
  \citenamefont {Hellerstedt}, \citenamefont {Edmonds}, \citenamefont
  {Kaczorowski}, \citenamefont {Hosen}, \citenamefont {Neupane}, \citenamefont
  {Lin}, \citenamefont {Fuhrer}, \citenamefont {Weber},\ and\ \citenamefont
  {Ishigami}}]{Lodge2017}%
  \BibitemOpen
  \bibfield  {author} {\bibinfo {author} {\bibfnamefont {M.~S.}\ \bibnamefont
  {Lodge}}, \bibinfo {author} {\bibfnamefont {G.}~\bibnamefont {Chang}},
  \bibinfo {author} {\bibfnamefont {C.-Y.}\ \bibnamefont {Huang}}, \bibinfo
  {author} {\bibfnamefont {B.}~\bibnamefont {Singh}}, \bibinfo {author}
  {\bibfnamefont {J.}~\bibnamefont {Hellerstedt}}, \bibinfo {author}
  {\bibfnamefont {M.~T.}\ \bibnamefont {Edmonds}}, \bibinfo {author}
  {\bibfnamefont {D.}~\bibnamefont {Kaczorowski}}, \bibinfo {author}
  {\bibfnamefont {M.~M.}\ \bibnamefont {Hosen}}, \bibinfo {author}
  {\bibfnamefont {M.}~\bibnamefont {Neupane}}, \bibinfo {author} {\bibfnamefont
  {H.}~\bibnamefont {Lin}}, \bibinfo {author} {\bibfnamefont {M.~S.}\
  \bibnamefont {Fuhrer}}, \bibinfo {author} {\bibfnamefont {B.}~\bibnamefont
  {Weber}}, \ and\ \bibinfo {author} {\bibfnamefont {M.}~\bibnamefont
  {Ishigami}},\ }\href {\doibase 10.1021/acs.nanolett.7b02307} {\bibfield
  {journal} {\bibinfo  {journal} {Nano Lett.}\ }\textbf {\bibinfo {volume}
  {17}},\ \bibinfo {pages} {7213} (\bibinfo {year} {2017})}\BibitemShut
  {NoStop}%
\bibitem [{\citenamefont {Nedoliuk}\ \emph {et~al.}(2019)\citenamefont
  {Nedoliuk}, \citenamefont {Hu}, \citenamefont {Geim},\ and\ \citenamefont
  {Kuzmenko}}]{Nedoliuk2019}%
  \BibitemOpen
  \bibfield  {author} {\bibinfo {author} {\bibfnamefont {I.~O.}\ \bibnamefont
  {Nedoliuk}}, \bibinfo {author} {\bibfnamefont {S.}~\bibnamefont {Hu}},
  \bibinfo {author} {\bibfnamefont {A.~K.}\ \bibnamefont {Geim}}, \ and\
  \bibinfo {author} {\bibfnamefont {A.~B.}\ \bibnamefont {Kuzmenko}},\ }\href
  {https://www.nature.com/articles/s41565-019-0489-8} {\bibfield  {journal}
  {\bibinfo  {journal} {Nat. Nanotechnol.}\ }\textbf {\bibinfo {volume} {14}},\
  \bibinfo {pages} {756} (\bibinfo {year} {2019})}\BibitemShut {NoStop}%
\bibitem [{\citenamefont {Topp}\ \emph {et~al.}(2017)\citenamefont {Topp},
  \citenamefont {Queiroz}, \citenamefont {Gr\"uneis}, \citenamefont
  {M\"uchler}, \citenamefont {Rost}, \citenamefont {Varykhalov}, \citenamefont
  {Marchenko}, \citenamefont {Krivenkov}, \citenamefont {Rodolakis},
  \citenamefont {McChesney}, \citenamefont {Lotsch}, \citenamefont {Schoop},\
  and\ \citenamefont {Ast}}]{Topp2017}%
  \BibitemOpen
  \bibfield  {author} {\bibinfo {author} {\bibfnamefont {A.}~\bibnamefont
  {Topp}}, \bibinfo {author} {\bibfnamefont {R.}~\bibnamefont {Queiroz}},
  \bibinfo {author} {\bibfnamefont {A.}~\bibnamefont {Gr\"uneis}}, \bibinfo
  {author} {\bibfnamefont {L.}~\bibnamefont {M\"uchler}}, \bibinfo {author}
  {\bibfnamefont {A.~W.}\ \bibnamefont {Rost}}, \bibinfo {author}
  {\bibfnamefont {A.}~\bibnamefont {Varykhalov}}, \bibinfo {author}
  {\bibfnamefont {D.}~\bibnamefont {Marchenko}}, \bibinfo {author}
  {\bibfnamefont {M.}~\bibnamefont {Krivenkov}}, \bibinfo {author}
  {\bibfnamefont {F.}~\bibnamefont {Rodolakis}}, \bibinfo {author}
  {\bibfnamefont {J.~L.}\ \bibnamefont {McChesney}}, \bibinfo {author}
  {\bibfnamefont {B.~V.}\ \bibnamefont {Lotsch}}, \bibinfo {author}
  {\bibfnamefont {L.~M.}\ \bibnamefont {Schoop}}, \ and\ \bibinfo {author}
  {\bibfnamefont {C.~R.}\ \bibnamefont {Ast}},\ }\href {\doibase
  10.1103/PhysRevX.7.041073} {\bibfield  {journal} {\bibinfo  {journal} {Phys.
  Rev. X}\ }\textbf {\bibinfo {volume} {7}},\ \bibinfo {pages} {041073}
  (\bibinfo {year} {2017})}\BibitemShut {NoStop}%
\bibitem [{\citenamefont {Duan}\ \emph {et~al.}(2019)\citenamefont {Duan},
  \citenamefont {Yang}, \citenamefont {Ma}, \citenamefont {Zhu},\ and\
  \citenamefont {Zhang}}]{Duan2019}%
  \BibitemOpen
  \bibfield  {author} {\bibinfo {author} {\bibfnamefont {W.}~\bibnamefont
  {Duan}}, \bibinfo {author} {\bibfnamefont {C.}~\bibnamefont {Yang}}, \bibinfo
  {author} {\bibfnamefont {Z.}~\bibnamefont {Ma}}, \bibinfo {author}
  {\bibfnamefont {Y.}~\bibnamefont {Zhu}}, \ and\ \bibinfo {author}
  {\bibfnamefont {C.}~\bibnamefont {Zhang}},\ }\href {\doibase
  10.1103/PhysRevB.99.045124} {\bibfield  {journal} {\bibinfo  {journal} {Phys.
  Rev. B}\ }\textbf {\bibinfo {volume} {99}},\ \bibinfo {pages} {045124}
  (\bibinfo {year} {2019})}\BibitemShut {NoStop}%
\end{thebibliography}%

\end{document}